\documentclass[sigconf]{acmart}

\usepackage{booktabs} % For formal tables
\usepackage{amsmath,amsfonts,amssymb}
\usepackage{eqnarray}
\usepackage{mathtools}
\usepackage{balance} 
\usepackage{mathrsfs}
\usepackage{multirow}
\usepackage{csquotes}
\usepackage{bbm}
\usepackage{enumitem} 
\usepackage{amsmath}
\usepackage{colortbl}
\usepackage[]{algorithm2e}
\usepackage{graphics}
\usepackage{subfigure}
\usepackage{units}
\usepackage{csquotes}
\usepackage{todonotes}

\makeatletter
\def\@fnsymbol#1{\ensuremath{\ifcase#1\or *\or \dagger\or \ddagger\or
   \mathsection\or \mathparagraph\or \|\or **\or \dagger\dagger
   \or \ddagger\ddagger \else\@ctrerr\fi}}
\makeatother

\newcommand{\greyrule}{\arrayrulecolor{black!30}\midrule\arrayrulecolor{black}}

\makeatletter
\renewcommand*\env@matrix[1][*\c@MaxMatrixCols c]{%
  \hskip -\arraycolsep
  \let\@ifnextchar\new@ifnextchar
  \array{#1}}
\makeatother

\copyrightyear{2019}
\acmYear{2019}
\setcopyright{acmcopyright}
\acmConference[SIGIR '19]{Proceedings of the 42nd International ACM SIGIR Conference on Research and Development in Information Retrieval}{July 21--25, 2019}{Paris, France}
\acmBooktitle{Proceedings of the 42nd International ACM SIGIR Conference on Research and Development in Information Retrieval (SIGIR '19), July 21--25, 2019, Paris, France}
\acmPrice{15.00}
\acmDOI{10.1145/3331184.3331316}
\acmISBN{978-1-4503-6172-9/19/07}
\begin{document}

% TODO title
% or? Improving Weak Supervision of Neural Information Retrieval Models by Modeling Query-Document Interactions
% ? Enabling the use of content-based weak supervision through interaction filtering [needs to mention NIR though?]
% ? Filtering Interactions to Improve Content-Based Weak Supervision for Neural Information Retrieval Models
% ? Improving Content-Based Weak Supervision for Neural Models through Interaction Filtering
\title{Content-Based Weak Supervision for Ad-Hoc Re-Ranking}

\author{Sean MacAvaney}
\affiliation{%
  \institution{IRLab, Georgetown University}
}
\email{sean@ir.cs.georgetown.edu}

\author{Andrew Yates}
\affiliation{%
  \institution{Max Planck Institute for Informatics}
}
\email{ayates@mpi-inf.mpg.de}

\author{Kai Hui}
\authornote{Work conducted while the author was at the Max Planck Institute for Informatics.}
\affiliation{%
  \institution{Amazon}
}
\email{kaihuibj@amazon.com}

\author{Ophir Frieder}
\affiliation{%
  \institution{IRLab, Georgetown University}
}
\email{ophir@ir.cs.georgetown.edu}

% The default list of authors is too long for headers.
% \renewcommand{\shortauthors}{S. MacAvaney et al.}

\begin{abstract}
One challenge with neural ranking is the need for a large amount of manually-labeled relevance judgments for training. In contrast with prior work, we examine the use of weak supervision sources for training that yield pseudo query-document \textit{pairs} that already exhibit relevance (e.g., newswire headline-content pairs and encyclopedic heading-paragraph pairs). We also propose filtering techniques to eliminate training samples that are too far out of domain using two techniques: a heuristic-based approach and novel supervised filter that re-purposes a neural ranker. Using several leading neural ranking architectures and multiple weak supervision datasets, we show that these sources of training pairs are effective on their own (outperforming prior weak supervision techniques), and that filtering can further improve performance.
\end{abstract}

%
% The code below should be generated by the tool at
% http://dl.acm.org/ccs.cfm
% Please copy and paste the code instead of the example below. 
%
%\begin{CCSXML}
%<ccs2012>
%<concept>
%<concept_id>10002951.10003317</concept_id>
%<concept_desc>Information systems~Information retrieval</concept_desc>
%<concept_significance>500</concept_significance>
%</concept>
%<concept>
%<concept_id>10002951.10003317.10003338.10003342</concept_id>
%<concept_desc>Information systems~Similarity measures</concept_desc>
%<concept_significance>500</concept_significance>
%</concept>
%<concept>
%<concept_id>10010147.10010257.10010258.10010260</concept_id>
%<concept_desc>Computing methodologies~Unsupervised learning</concept_desc>
%<concept_significance>500</concept_significance>
%</concept>
%</ccs2012>
%\end{CCSXML}

%\ccsdesc[500]{Information systems~Information retrieval}
%\ccsdesc[500]{Information systems~Similarity measures}
%\ccsdesc[500]{Computing methodologies~Unsupervised learning}

%\keywords{, Weak supervision}

\maketitle

% Introduction
\section{Introduction} 
\label{sec.introduction}

A lack of manual training data is a perennial problem in information retrieval~\cite{Zamani2018SIGIR2W}.
To enable training supervised rankers for new domains, we propose a weak supervision approach based on \textit{pairs} of text to train neural ranking models and a filtering technique to adapt the dataset to a given domain. Our approach eliminates the need for a query log or large amounts of manually-labeled in-domain relevance judgments to train neural rankers, and exhibits stronger and more varied positive relevance signals than prior weak supervision work (which relies on BM25 for these signals).

Others have experimented with weak supervision for neural ranking (see Section~\ref{sec.background.weaksupervision}). Our weak supervision approach differs from these approaches in a crucial way: we train neural rankers using datasets of text \textit{pairs} that exhibit relevance, rather than using a heuristic to find pseudo-relevant documents for queries. For instance, the text pair from a newswire dataset consisting of an article's headline and its content exhibits an inherent sense of relevance because a headline often provides a concise representation of an article's content. To overcome possible domain differences between the training data and the target domain,
we propose an approach to filter the training data using a small set of queries (templates) from the target domain. We evaluate two filters: an unsupervised heuristic and using the neural ranker itself as a discriminator.

We evaluate our approaches by training several leading neural ranking architectures on two sources of weak supervision text pairs. We show that our methods can significantly outperform various neural rankers when trained using a query log source (as proposed by~\cite{dehghani2017neural}), the ranker when trained on a limited amount of manually-labeled in-domain data (as one would encounter in a new domain), and well-tuned conventional baselines. 
In summary, we (1) address existing shortcomings of weak supervision to train neural rankers by using training sources from text pairs, (2) address limitations related to domain differences when training rankers on these sources using novel filtering techniques, and (3) demonstrate the effectiveness of our methods for ad-hoc retrieval when limited in-domain training data is available. Our code is public for validation and further comparisons.\footnote{\url{https://github.com/Georgetown-IR-Lab/neuir-weak-supervision}}

% Background
\section{Background and related work}
\label{sec.background}
\subsection{Neural IR models}\label{sec.background.nir}
Ad-hoc retrieval systems rank documents according to their
relevance to a given query.
A neural IR model
($\mathit{nir}$) aims to measure the interaction between a query-document pair $(q,
\,d)$ with a real-value relevance score $rel=\mathit{nir}(q,d)$. The model $nir$ is trained to minimize pairwise loss between training triples consisting of a query $q$, relevant document $d^+$, and non-relevant document $d^-$.
Neural retrieval models can be categorized as \emph{semantic matching} models (which create dense query/document representations) or as \emph{relevance matching} models (which compare query and document terms directly, often through a query-document similarity matrix). We focus on relevance matching models because they generally show better performance than semantic matching models.
We test our approach on three leading neural rankers:

\textbf{KNRM}~\cite{xiong2017end} uses Gaussian kernels applied to each individual similarity score and log-summed across the document dimension. A final dense learning-to-rank phase combines these features into a relevance score.

\textbf{Conv-KNRM}~\cite{convknrm} is a variant of KNRM which applies convolution filters of lengths 1--3 over word embeddings before building cross-matched (matching all kernel lengths with one another) similarity matrices. The rest of the ranking process is identical to KNRM.

\textbf{PACRR}~\cite{hui2017pacrr}
uses square convolutional kernels over the similarity matrix to capture soft n-gram matches. $k-$max pooling is applied to retain only the strongest signals for each query term, and signals are combined with a dense layer.

\subsection{Weak supervision}\label{sec.background.weaksupervision}
In IR, weak supervision uses pseudo-relevant information to train a ranking model in place of human judgments. Early work on weak supervision for IR focused on training learning-to-rank models~\cite{azzopardi2007building},
using web anchor text~\cite{asadi2011pseudo} and microblog hashtags~\cite{berendsen2013pseudo} for weak supervision. More recently, \citet{dehghani2017neural} proposed a weak supervision approach that makes use of the AOL query log and BM25 results as a source of training data. Aside from limitations surrounding the availability of query logs, their approach suffers from limitations of BM25 itself: it assumes that documents ranked higher by BM25 are more relevant to the query than documents ranked lower. Others have suggested using a similar approach, but using news headlines~\cite{Li2018JointLF}, also assuming relevance from BM25 rankings. Still others have employed a Generative Adversarial Network to build training samples~\cite{Wang2017IRGANAM}, but this limits the generated data to the types of relevance found in the training samples, making it a complementary approach. In contrast, our approach uses freely-available text \textit{pairs} that exhibit both a high quality and large size.

% Method
\section{Method} 
\label{sec.method}

\subsection{Ranking- and content-based sources}\label{sec.source}

Recall that pairwise training consists of a set of training triples, each consisting of a query $q$, relevant document $d^+$, and non-relevant document $d^-$. We describe two sources of weak supervision training data that replace human-generated relevance judgments: ranking-based and content-based training sources.

\textbf{Ranking-based training sources}, first proposed by \cite{dehghani2017neural}, are defined by a collection of texts $T$, a collection of documents $D$, and an unsupervised ranking function $R(q,d)$ (e.g., BM25). Training triples are generated as follows. Each text is treated as a query $q\in T$. All documents in $D$ are ranked using $R(\cdot)$, giving $D^q$. Relevant documents are sampled using a cutoff $c^+$, and non-relevant documents are sampled using cutoff $c^-$, such that $d^+\in D^q[0:c^+]$ and $d^-\in D^q[c^+:c^-]$. This source is referred to as ranking-based because the unsupervised ranker is the source of relevance.\footnote{Our formulation of ranking-based sources is slightly different than what was proposed by \citet{dehghani2017neural}: we use cutoff thresholds for positive and negative training samples, whereas they suggest using random pairs. Pilot studies we conducted showed that the threshold technique usually performs better.}

\textbf{Content-based training sources} are defined as a collection of text pairs $P=\{(a_1,b_1),(a_2,b_2),...,(a_{|P|},b_{|P|})\}$ and an unsupervised ranking function $R(q,d)$ (e.g., BM25). The text pairs should be semantically related pairs of text, where the first element is similar to a query, and the second element is similar to a document in the target domain. For instance, they could be heading-content pairs of news articles (the headline describes the content of the article content). For a given text pair, a query and relevant document are selected $(q,d^+)\in P$. The non-relevant document is selected from the collection of documents in $B=\{b_1,b_2,...,b_{|P|}\}$. We employ $R(\cdot)$ to select challenging negative samples from $B^q$. A negative cutoff $c^-$ is employed, yielding negative document $d^-\in B^q[0:c^-]-\{d^+\}$. We discard positive samples where $d^+$ is not within this range to eliminate overtly non-relevant documents. This approach can yield documents relevant to $q$, but we assert that $d^+$ is \textit{more} relevant.

Although ranking-based and content-based training sources bear some similarities, important differences remain. Content-based sources use text pairs as a source of positive relevance, whereas ranking-based sources use the unsupervised ranking. Furthermore, content-based sources use documents from the pair's domain, not the target domain. We hypothesize that the enhanced notion of relevance that content-based sources gain from text pairs will improve ranking performance across domains, and show this in Section~\ref{sec.evaluation}.

\subsection{Filter framework}\label{sec.method.filters}
We propose a filtering framework to overcome domain mismatch that can exist between data found in a weak supervision training source and data found in the target dataset. The framework consists of a filter function $F_D(q,d)$ that determines the suitability of a given weak supervision query-document pair $(q,d)$ to the domain $D$. All relevant training pairs $(q,d^+)\in S$ for a weak supervision source $S$ are ranked using $F_D(q,d^+)$ and the $c_{max}$ maximum pairs are chosen: $
S_D=\max^{c_{max}}_{(q,d^+)\in S}F_D(q,d^+)
$. To tune $F_D(\cdot)$ to domain $D$, a set of \textit{template pairs} from the target domain are employed. The set of pairs $T_D$ % =\{(q^D_1,d^D_1),(q^D_2,d^D_2),...,(q^D_{|T_D|},d^D_{|T_D|})\}
is assumed to be relevant in the given domain.\footnote{Templates do not require human judgments. We use sample queries and an unsupervised ranker to generate $T_D$. Manual judgments can be used when available.} We assert that these filters are easy to design and can have broad coverage of ranking architectures. We present two implementations of the filter framework: the $k$max filter, and the Discriminator filter.

\textbf{$k$-Maximum Similarity ($k$max) filter.}
This heuristic-based filter consists of two components: a \textit{representation function} $\mathit{rep}(q,d)$ and a \textit{distance function} $\mathit{dist}(r_1,r_2)$.
The representation function captures some matching signal between query $q$ and document $d$ as a vector. Since many neural ranking models consider similarity scores between terms in the query and document to perform soft term matching~\cite{hui2017pacrr,convknrm,xiong2017end,guo2016deep}, this filter selects the $k$ maximum cosine similarity scores between the word vectors of each query term and all terms in the document:
$\max_{d_j\in d}^ksim(q_i, d_j) : \forall q_i \in q$.

Since neural models can capture local patterns (e.g., n-grams), we use an aligned mean square error. The aligned MSE iterates over possible configurations of elements in the representation by shifting the position to find the alignment that yields the smallest distance. In other words, it represents the minimum mean squared error given all rotated configurations of the query.
Based on the shift operation and given two interaction representation matrices $r_1$ and $r_2$,
the aligned $\mathit{dist}_{kmax}(r_1,r_2)$ is defined as 
the minimum distance when shifting $r_1$ for $s\in [1, |r_1|)$.
More formally: $\mathit{dist}_{kmax}(r_1,r_2) = \min_{s=1}^{|r_1|}{\mathit{MSE}\big(\mathit{shift}(r_1, s),r_2\big)}$.

Using these two functions, the filter is simply defined as the minimum distance between the representations of it and any template pair from the target domain:

\vspace{-1.2em}\begin{equation}\label{eq.dist.drmm}
F_D(q,d)=\min_{(q',d')\in T_D}dist(rep(q,d),rep(q',d'))
\end{equation}\vspace{-0.8em}

\textbf{Discriminator filter.}
A second approach to interaction filtering is to use the ranking architecture $R$ itself. Rather than training $R$ to distinguish different degrees of relevance, here we use $R$ to train a model to distinguish between samples found in the weak supervision source and $T_D$. This technique employs the same pairwise loss approach used for relevance training and is akin to the discriminator found in generative adversarial networks. Pairs are sampled uniformly from both templates and the weak supervision source. Once $R_D$ is trained, all weak supervision training samples are ranked with this model acting as $F_D(\cdot)=R_D(\cdot)$.

The intuition behind this approach is that the model should learn characteristics that distinguish in-domain pairs from out-of-domain pairs, but it will have difficulty distinguishing between cases where the two are similar. One advantage of this approach is that it allows for training an interaction filter for any arbitrary ranking architecture, although it requires a sufficiently large $T_D$ to avoid overfitting.

% % 
%Evaluation
\section{Evaluation} 
\label{sec.evaluation}

\subsection{Experimental setup}\label{sec.expsetting}

\textbf{Training sources.} We use the following four sources of training data to verify the effectiveness of our methods:
\begin{itemize}[leftmargin=*]
\item[-] \textbf{Query Log (AOL, ranking-based, $100k$ queries).}
This source uses the AOL query log~\cite{pass2006} as the basis for a ranking-based source, following the approach of~\cite{dehghani2017neural}.\footnote{
Distinct non-navigational queries from the AOL query log from March 1, 2006 to May 31, 2006 are selected. We randomly sample $100k$ of queries with length of at least 4. While \citeauthor{dehghani2017neural}~\cite{dehghani2017neural} used a larger number of queries to train their model, the state-of-the-art relevance matching models we evaluate do not learn term embeddings (as \cite{dehghani2017neural} does) and thus converge with fewer than $100k$ training samples.}
We retrieve ClueWeb09 documents for each query using the Indri\footnote{https://www.lemurproject.org/indri/} query likelihood~(QL) model. We fix $c^+=1$ and $c^-=10$ due to the expense of sampling documents from ClueWeb.

\item[-] \textbf{Newswire (NYT, content-based, $1.8m$ pairs).} 
We use the New York Times corpus~\cite{sandhaus2008new} as a content-based source, using headlines as pseudo queries and the corresponding content as pseudo relevant documents. We use BM25 to select the negative articles, retaining top $c^-=100$ articles for individual headlines.

\item[-] 
\textbf{Wikipedia (Wiki, content-based, $1.1m$ pairs).} 
Wikipedia article heading hierarchies and their corresponding paragraphs have been employed as a training set for the \textsc{Trec} Complex Answer Retrieval (CAR) task~\cite{Nanni2017BenchmarkFC,macavaney2018overcoming}.
We use these pairs as a content-based source, assuming that the hierarchy of headings is a relevant query for the paragraphs under the given heading.
Heading-paragraph pairs from train fold 1 of the \textsc{Trec} CAR dataset~\cite{Dietz2017} (v1.5) are used. We generate negative heading-paragraph pairs for each heading using BM25 ($c^-=100$).

\item[-] \textbf{Manual relevance judgments (WT10).}
We compare the ranking-based and content-based sources with a data source that consists of relevance judgments generated by human assessors. In particular, manual judgments from 2010 \textsc{Trec} Web Track ad-hoc task (WT10) are employed, which includes $25k$ manual relevance judgments ($5.2k$ relevant) for 50 queries (topics + descriptions, in line with~\cite{hui2017pacrr,guo2016deep}). This setting represents a new target domain, with limited (yet still substantial) manually-labeled data.

\end{itemize}

\textbf{Training neural IR models.}
We test our method using several state-of-the-art neural IR models (introduced in Section~\ref{sec.background.nir}):
PACRR~\cite{hui2017pacrr},
Conv-KNRM~\cite{convknrm}, and
KNRM~\cite{xiong2017end}.\footnote{By using these stat-of-the-art architectures, we are using stronger baselines than those used in~\cite{dehghani2017neural,Li2018JointLF}.}
We use the model architectures and hyper-parameters (e.g., kernel sizes) from the best-performing configurations presented in the original papers for all models.
All models are trained using pairwise loss for 200 iterations with 512 training samples each iteration.
We use Web Track 2011 (WT11) manual relevance judgments as validation data to select the best iteration via nDCG@20. This acts as a way of fine-tuning the model to the particular domain, and is the only place that manual relevance judgments are used during the weak supervision training process. At test time, we re-rank the top 100 Indri QL results for each query.

\textbf{Interaction filters.}
We use the 2-maximum and discriminator filters for each ranking architecture to evaluate the effectiveness of the interaction filters.
We use queries from the target domain (\textsc{Trec} Web Track 2009--14) to generate the template pair set for the target domain $T_D$.
To generate pairs for $T_D$, the top 20 results from query likelihood (QL) for individual queries on ClueWeb09 and ClueWeb12\footnote{\url{https://lemurproject.org/clueweb09.php}, \url{https://lemurproject.org/clueweb12.php}} are used to construct query-document pairs.
Note that this approach makes no use of manual relevance judgments because only query-document pairs from the QL search results are used (without regard for relevance).
We do not use query-document pairs from the target year to avoid any latent query signals from the test set. The supervised discriminator filter is validated using a held-out set of 1000 pairs. To prevent overfitting the training data, we reduce the convolutional filter sizes of PACRR and ConvKNRM to 4 and 32, respectively. We tune $c_{max}$ with the validation dataset (WT11) for each model ($100k$ to $900k$, $100k$ intervals).

\textbf{Baselines and benchmarks.}
As baselines, we use the AOL ranking-based source as a weakly supervised baseline~\cite{dehghani2017neural}, WT10 as a manual relevance judgment baseline, and BM25 as an unsupervised baseline. The two supervised baselines are trained using the same conditions as our approach, and the BM25 baselines is tuned on each testing set with Anserini~\cite{Yang2017AnseriniET}, representing the best-case performance of BM25.\footnote{Grid search: $b\in[0.05,1]$ (0.05 interval), and $k_1\in[0.2,4]$ (0.2 interval)}
We measure the performance of the models using
the \textsc{Trec} Web Track 2012--2014 (WT12--14) queries (topics + descriptions) and manual relevance judgments. These cover two target collections: ClueWeb09 and ClueWeb12.
Akin to~\cite{dehghani2017neural}, the trained models are used to
re-rank the top 100 results from a query-likelihood model (QL, Indri~\cite{strohman2005indri} version).
Following the \textsc{Trec} Web Track, we use
nDCG@20 and ERR@20 for evaluation.

\begin{table}
\scriptsize
\caption{Ranking performance when trained using content-based sources (NYT and Wiki). Significant differences compared to the baselines ([B]M25, [W]T10, [A]OL) are indicated with $\uparrow$ and $\downarrow$ (paired t-test, $p<0.05$).}\label{tab.results}
\vspace{-1em}
\begin{tabular}{llrrrrrr}
\toprule
&&\multicolumn{3}{c}{nDCG@20} \\\cmidrule(lr){3-5}
        Model &   Training & WT12 & WT13 & WT14 \\
\midrule

\multicolumn{2}{l}{BM25 (tuned w/~\cite{Yang2017AnseriniET})} & 0.1087  & 0.2176  & 0.2646 \\
\midrule
PACRR & WT10 & B$\uparrow$ 0.1628  & 0.2513  & 0.2676 \\
 & AOL & 0.1910  & 0.2608  & 0.2802 \\
\greyrule
 & NYT & \bf W$\uparrow$ B$\uparrow$ 0.2135  & \bf A$\uparrow$ W$\uparrow$ B$\uparrow$ 0.2919  & \bf W$\uparrow$ 0.3016 \\
 & Wiki & W$\uparrow$ B$\uparrow$ 0.1955  & A$\uparrow$ B$\uparrow$ 0.2881  & W$\uparrow$ 0.3002 \\
\midrule
Conv-KNRM & WT10 & B$\uparrow$ 0.1580  & 0.2398  & B$\uparrow$ 0.3197 \\
 & AOL & 0.1498  & 0.2155  & 0.2889 \\
\greyrule
 & NYT & \bf A$\uparrow$ B$\uparrow$ 0.1792  & \bf A$\uparrow$ W$\uparrow$ B$\uparrow$ 0.2904  & \bf B$\uparrow$ 0.3215 \\
 & Wiki & 0.1536  & A$\uparrow$ 0.2680  & B$\uparrow$ 0.3206 \\
\midrule
KNRM & WT10 & B$\uparrow$ 0.1764  & \bf 0.2671  & 0.2961 \\
 & AOL & \bf B$\uparrow$ 0.1782  & 0.2648  & \bf 0.2998 \\
\greyrule
 & NYT & W$\downarrow$ 0.1455  & A$\downarrow$ 0.2340  & 0.2865 \\
 & Wiki & A$\downarrow$ W$\downarrow$ 0.1417  & 0.2409  & 0.2959 \\

\bottomrule
\end{tabular}
\vspace{-2em}
\end{table}

\renewcommand{\arraystretch}{1}

\subsection{Results}\label{sec.results}
In Table~\ref{tab.results}, we present the performance of the rankers when trained using content-based sources without filtering.
In terms of absolute score, we observe that the two n-gram models (PACRR and ConvKNRM) always perform better when trained on content-based sources than when trained on the limited sample of in-domain data. When trained on NYT, PACRR performs significantly better. KNRM performs worse when trained using the content-based sources, sometimes significantly. These results suggest that these content-based training sources contain relevance signals where n-grams are useful, and it is valuable for these models to see a wide variety of n-gram relevance signals when training. The n-gram models also often perform significantly better than the ranking-based AOL query log baseline. This makes sense because BM25's rankings do not consider term position, and thus cannot capture this important indicator of relevance. This provides further evidence that content-based sources do a better job providing samples that include various notions of relevance than ranking-based sources.

When comparing the performance of the content-based training sources, we observe that the NYT source usually performs better than Wiki. We suspect that this is due to the web domain being more similar to the newswire domain than the complex answer retrieval domain. For instance, the document lengths of news articles are more similar to web documents, and precise term matches are less common in the complex answer retrieval domain~\cite{macavaney2018overcoming}.

We present filtering performance on NYT and Wiki for each ranking architecture in Table~\ref{tab:filter_results}. In terms of absolute score, the filters almost always improve the content-based data sources, and in many cases this difference is statistically significant. The one exception is for Conv-KNRM on NYT. One possible explanation is that the filters caused the training data to become too homogeneous, reducing the ranker's ability to generalize. We suspect that Conv-KNRM is particularly susceptible to this problem because of language-dependent convolutional filters; the other two models rely only on term similarity scores. We note that Wiki tends to do better with the 2max filter, with significant improvements seen for Conv-KNRM and KNRM. In thse models, the discriminator filter may be learning surface characteristics of the dataset,
rather than more valuable notions of relevance. We also note that $c_{max}$ is an important (yet easy) hyper-parameter to tune, as the optimal value varies considerably between systems and datasets.

%Related Work
%\input{Relatedwork}

% %  
%Conclusion
\section{Conclusion} 
\label{sec.conclusion}

We presented an approach for employing content-based sources of pseudo relevance for training neural IR models. We demonstrated that our approach can match (and even outperform) neural ranking models trained on manual relevance judgments and existing ranking-based weak supervision approaches using two different sources of data. We also showed that performance can be boosted using two filtering techniques: one heuristic-based and one that re-purposes a neural ranker. By using our approach, one can effectively train neural ranking models on new domains without behavioral data and with only limited in-domain data.

\begin{table}
\centering
\scriptsize
\caption{Ranking performance using filtered NYT and Wiki. Significant improvements and reductions compared to unfiltered dataset are marked with $\uparrow$ and $\downarrow$ (paired t-test, $p<0.05$).}
\vspace{-1em}
\label{tab:filter_results}
\begin{tabular}{llrrrr}
\toprule
&&&\multicolumn{2}{c}{WebTrack 2012--14} \\\cmidrule(lr){4-5}
Model & Training & $k_{max}$ & nDCG@20 & ERR@20 \\
\midrule

PACRR &  NYT &  & 0.2690  & 0.2136 \\
 & w/ 2max & $200k$ & 0.2716  & 0.2195 \\
 & w/ discriminator & $500k$ & \bf $\uparrow$ 0.2875  & \bf 0.2273 \\
\midrule
 & Wiki &  & 0.2613  & 0.2038 \\
 & w/ 2max & $700k$ & 0.2568  & 0.2074 \\
 & w/ discriminator & $800k$ & \bf 0.2680  & \bf 0.2151 \\
\midrule
Conv-KNRM &  NYT &  & 0.2637  & 0.2031 \\
 & w/ 2max & $100k$ & $\downarrow$ 0.2338  & \bf 0.2153 \\
 & w/ discriminator & $800k$ & \bf 0.2697  & 0.1937 \\
\midrule
 & Wiki &  & 0.2474  & 0.1614 \\
 & w/ 2max & $400k$ & \bf 0.2609  & \bf $\uparrow$ 0.1828 \\
 & w/ discriminator & $700k$ & 0.2572  & 0.1753 \\
\midrule
KNRM &  NYT &  & 0.2220  & 0.1536 \\
 & w/ 2max & $100k$ & 0.2235  & \bf $\uparrow$ 0.1828 \\
 & w/ discriminator & $300k$ & \bf 0.2274  & $\uparrow$ 0.1671 \\
\midrule
 & Wiki &  & 0.2262  & 0.1635 \\
 & w/ 2max & $600k$ & \bf $\uparrow$ 0.2389  & \bf $\uparrow$ 0.1916 \\
 & w/ discriminator & $700k$ & 0.2366  & 0.1740 \\

\bottomrule
\end{tabular}
\vspace{-2em}
\end{table}

\bibliographystyle{ACM-Reference-Format}
\bibliography{main}

\end{document}